\documentclass[aps,prb,twocolumn,showpacs]{revtex4}
\usepackage{graphicx}
\usepackage{natbib}

\begin{document}
\title{Complex lithium ion dynamics in simulated LiPO$_3$ glass studied by means of multi-time correlation functions}

\author{Michael Vogel}
\affiliation{Department of Chemical Engineering, University of
Michigan, 2300 Hayward, Ann Arbor, MI, 48109, USA}

\date{\today}

\begin{abstract}
Molecular dynamics simulations are performed to study the lithium
jumps in LiPO$_3$ glass. In particular, we calculate higher-order
correlation functions that probe the positions of single lithium ions
at several times. Three-time correlation functions show that the
non-exponential relaxation of the lithium ions results from both
correlated back-and-forth jumps and the existence of dynamical
heterogeneities, i.e., the presence of a broad distribution of jump
rates. A quantitative analysis yields that the contribution of the
dynamical heterogeneities to the non-exponential depopulation of the
lithium sites increases upon cooling. Further, correlated
back-and-forth jumps between neighboring sites are observed for the
fast ions of the distribution, but not for the slow ions and, hence,
the back-jump probability depends on the dynamical state. Four-time
correlation functions indicate that an exchange between fast and slow
ions takes place on the timescale of the jumps themselves, i.e., the
dynamical heterogeneities are short-lived. Hence, sites featuring
fast and slow lithium dynamics, respectively, are intimately mixed.
In addition, a backward correlation beyond the first neighbor shell
for highly mobile ions and the presence of long-range dynamical
heterogeneities suggest that fast ion migration occurs along
preferential pathways in the glassy matrix. In the melt, we find no
evidence for correlated back-and-forth motions and dynamical
heterogeneities on the length scale of the next-neighbor distance.
\end{abstract}

\maketitle

\section{Introduction}

In view of numerous technological applications and a large variety
of phenomena, recent scientific work repeatedly focused on glassy
ion conductors. On the one hand, it is well known that the
macroscopic charge transport in these materials results from the
diffusion of mobile ions in a basically rigid glassy matrix, on
the other, a detailed microscopic picture of this dynamical
process is still lacking. One key feature of the dynamics in
glassy ion conductors is the non-exponential relaxation of the
mobile ions, indicating the complexity of the motion. For
instance, broad loss peaks are found in dielectrical and
mechanical relaxation experiments
\cite{Moynihan,Angell,Sidebottom,Green1,Green2}. Moreover, when
the dynamics of the mobile ions are studied by means of advanced
nuclear magnetic resonance (NMR) techniques, non-exponential
correlation functions are observed
\cite{Kreta,Bunsen,Spiess,Boehmer}. In general, the origin of
non-exponential relaxation can be twofold \cite{12}. In the
homogeneous scenario, the dynamics of all particles are
characterized by the same relaxation function that is
intrinsically non-exponential due to a specific mechanism of the
motion, e.g., correlated back-and-forth jumps. Contrarily, in the
heterogeneous scenario, all particles relax exponentially, i.e.,
there are no correlated back-and-forth jumps, but a distribution
of correlation times $G(\lg \tau)$ exists. In our case of a
complex motion, one may expect that both homogeneous and
heterogeneous dynamics contribute to the non-exponential
relaxation.

A strong frequency dependence of the electrical conductivity
$\sigma(\omega)$ of ion conductors indicates that back-and-forth
motions occur during ionic diffusion
\cite{Jonscher,Martin,Ingram,Elliot,Ngai,Funke,Roling}. However,
to study a complex motion in detail, it is not sufficient to probe
the dynamics at two times, but one has to resort to multi-time
correlation functions \cite{12}. For example, three-time
correlation functions can be used to quantify the extent to which
the non-exponential relaxation results from the homogeneous and
the heterogeneous scenario, respectively \cite{Heuer}. Four-time
correlation functions allow one to measure the lifetime of
dynamical heterogeneities, i.e., the timescale of exchange
processes between fast and slow ions. Two methods have recently
proven well suited to record multi-time correlation functions for
glassy ion conductors. While ion dynamics on the
$\mathrm{ps\!-\!ns}$ timescale can be studied in molecular
dynamics (MD) simulations \cite{MD2}, multi-dimensional NMR
techniques probe jumps on a timescale of $\mathrm{ms\!-\!s}$.
Specifically, multi-dimensional $^{109}$Ag NMR experiments showed
that a broad rate distribution governs the jumps of silver ions in
silver phosphate based glasses \cite{Bunsen}. Hence, both
correlated back-and-forth jumps and dynamical heterogeneities
indeed contribute to the non-exponentiality. However, to the best
of our knowledge the relevance of both contributions has not yet
been quantified for glassy ion conductors.

MD simulations are an ideal tool to investigate dynamical
processes on a microscopic level. Prior work on the origin of the
non-exponential relaxation in glassy ion conductors has focused on
alkali silicate glasses. For lithium silicate glasses, it was
reported that dynamical heterogeneities contribute to the
non-exponential relaxation of the lithium ions where the lifetime
of these heterogeneities is limited \cite{MD2,Habasaki}. Further,
back-and-forth motions on various length scales were observed,
including a backward correlation beyond the first neighbor shell
at low temperatures \cite{MD2}. For sodium silicate glasses, no
evidence for back-and-forth jumps between adjacent sodium sites
was found at higher temperatures \cite{Kob}. Moreover, several
workers demonstrated that, though there is no micro-segregation,
the sodium ions follow preferential pathways in the glassy matrix
\cite{Jund,Horbach}. The sodium ions inside these channels show a
higher mobility than the ones outside and, hence, dynamical
heterogeneities exist \cite{Sunyer}. In addition, studying
examples of ionic trajectories the mechanism of alkali ion
migration in silicate glasses was described as ''vacancy-like
process'' \cite{Cormack}. During a correlated motion of two
cations, an alkali site is left by one ion an occupied by another.

In the present work, we use MD simulations to analyze the complex
mechanism of ion dynamics in an alkali phosphate glass. In this
way, we intend to complement the prior studies on silicate glasses
and to compare the nature of ion diffusion in different amorphous
materials. Moreover, based on an analysis of novel higher-order
correlation functions, new information about this dynamical
process is revealed. In detail, the dynamics of non-crystalline
LiPO$_3$ is investigated in a temperature range where the motions
of the various atomic species strongly decouple upon cooling.
While the lithium ionic subsystem can be equilibrated at all
chosen temperatures, the diffusion of the phosphorus and the
oxygen particles freezes in on the ns-timescale of the simulation.
In other words, the dynamics in the melt and in the glass are
studied. Our main goal is to identify the temperature dependent
origin of the non-exponential relaxation of the lithium ions. For
this purpose, we calculate multi-time correlation functions that
link the positions of single ions at subsequent points in time. In
this way, we quantify the homogeneous and the heterogeneous
contributions to the non-exponential relaxation and measure the
lifetime of the dynamical heterogeneities. Since the multi-time
correlation functions observed in NMR experiments on solid ion
conductors and some of the quantities computed in the present
study have a similar information content, a future comparison of
the respective results is very promising.

\section{Details of the simulation}

The potential used to describe the interaction of the ions in
LiPO$_3$ can be written as the sum of a Coulomb and a
Born-Mayer-Huggins pair potential
\begin{equation}\label{potential}
\Phi_{\alpha\beta}(r)=\frac{q_{\alpha}q_{\beta}\,e^2}{r}+
A_{\alpha\beta}\,\exp\left(-r/\rho\right)
\end{equation}
where $r$ is the distance between two ions of type $\alpha$ and
type $\beta$, respectively. The potential parameters, except for
$A_{LiO}$, are adopted from the work of Karthikeyan et al.\
\cite{Kart}. Here, $A_{LiO}$ is reduced to obtain a more realistic
interatomic distance $r_{LiO}$, cf.\ below. In detail, we use
effective charges $q_{Li}\!=\!0.6$, $q_{P}\!=\!3.0$ and
$q_{O}\!=\!-\!1.2$ as well as $\rho\!=\!0.29\mathrm \AA$. The
parameters $A_{\alpha\beta}$ are listed in Tab.\ 1. Most
simulations are performed in the NEV ensemble for $N\!=\!800$
particles. To take into account the thermal expansion of phosphate
glasses, the density is fixed at $\rho\!=\!2.15\,\mathrm{g/cm^3}$
as compared to $\rho\!=\!2.25\,\mathrm{g/cm^3}$ observed in
experiments at room temperature \cite{Muru}. This approximation
for the density at higher temperatures is obtained based on the
linear expansion coefficient of LiPO$_3$ glass \cite{English}. In
addition, we consider systems with $N\!=\!200$ and $N\!=\!400$ to
check for finite size effects. Moreover, some simulations are
performed at ''constant pressure'' to enable a comparison with
experimentally determined activation energies. Strictly speaking,
we continue doing runs in the NEV ensemble, but the size of the
simulation box is adjusted so that a pressure
$p\!=\!7.0\pm0.1\mathrm{GPa}$ is obtained at all temperatures.
This procedure results in a decrease of the density from
$\rho\!=\!2.21\,\mathrm{g/cm^3}$ at $744\mathrm K$ to
$\rho\!=\!1.95\,\mathrm{g/cm^3}$ at $3000\mathrm K$. The equations
of motion are integrated using the velocity form of the Verlet
algorithm with a time step of $2\mathrm{fs}$. Further, periodic
boundary conditions are applied and the Coulombic forces are
calculated via Ewald summation. In doing so, consistent results
are obtained when computing the trajectories with the programme
MOLDY \cite{Refson} and an own MD simulation programme,
respectively. While the whole system can be equilibrated at
$T\!\geq\!1006\mathrm K$, the phosphate glass matrix is basically
frozen on the timescale of the simulation at lower temperatures.
In the latter cases, we pay attention that the lithium ionic
subsystem is still in equilibrium. For example, the sample is
equilibrated for $20\mathrm{ns}$ at $T\!=\!592\mathrm K$ before
recording the data.

\begin{table}
\caption{Interaction parameters $A_{\alpha\beta}$, cf.\ Eq.\ 1,
together with the interatomic distances $r_{\alpha\beta}$ as obtained
from the present simulations and from experimental work
\cite{Muru,Beaufils}, respectively.}
\begin{math}
\begin{array}{|c|c|c|c|} \hline
& \mathrm {A_{\alpha\beta}\;[eV]} &  \mathrm{r_{\alpha\beta}\;[\AA]} & \mathrm{r_{\alpha\beta}\;[\AA]}\\
& &\mathrm{simulation}&\mathrm{experiment}\\
 \hline
\mathrm {LiLi} & 167.47 & 2.74 &  \\
\hline
\mathrm{LiP} & 158.09 & 3.31 &   \\
\hline
\mathrm{LiO} & 300.00 & 2.03 & 2.02 \\
\hline
\mathrm{PP} & 148.93 & 3.33 & 3.01\\
\hline
\mathrm{PO} & 694.66 & 1.51/1.66 & 1.52\!-\!1.56\\
\hline
\mathrm{OO} & 2644.8 & 2.58 & 2.52 \\
\hline
\end{array}
\end{math}
\end{table}

Karthikeyan et al.\ \cite{Kart} demonstrated that the structure of
LiPO$_3$ glass is well reproduced when using the two-body
potential specified in Eq.\ \ref{potential}. We confirmed this
result by analyzing the structure of the glass obtained after a
quench from $T\!=\!592\mathrm K$ to $T\!=\!300 \mathrm K$. Tab.\ 1
shows that the interatomic distances $r_{\alpha\beta}$ are in good
agreement with the corresponding experimental values
\cite{Muru,Beaufils}. Typical of phosphate glasses \cite{Hoppe},
the P-O distance for tangling oxygens
($r_{PTO}\!=\!1.51\mathrm\AA$) is shorter than for bridging
oxygens ($r_{PBO}\!=\!1.66\mathrm\AA$). Further, the reduction of
$A_{LiO}$, as compared to the value used by Karthikeyan et al.\
\cite{Kart}, improves the agreement with the actual distance
$r_{LiO}$ \cite{Muru,Beaufils}. Consistent with experimental
findings for LiPO$_3$ glass \cite{Hoppe,Brow,Wullen}, the
bond-angle distributions and the coordination numbers indicate
that the simulated glass consists of well defined phosphate
tetrahedra that are connected by two of their corners to form long
chains and/or rings. Concerning the intermediate range order, some
differences between simulated and actual LiPO$_3$ glass may exist.
In agreement with results by Karthikeyan et al.\ \cite{Kart}, we
find a mean P-O-P bond angle that is smaller than the one observed
in experiments on vitreous P$_2$O$_5$ \cite{Hoppe}. To fix this
mean angle at the experimental value for vitreous P$_2$O$_5$,
Liang et al.\ \cite{Alam} used a three-body potential in
simulations of LiPO$_3$ glass. We refrain from doing so because,
first, the P-O-P bond angle in meta-phosphate glasses has not yet
been determined experimentally and, second, three-body
interactions distinctly slow down the simulation. For our
analysis, it is important to perform simulations at sufficiently
low temperatures, because not until then lithium dynamics show
several characteristic features, e.g., back-and-forth jumps
between adjacent sites \cite{MD2,MD1}. Therefore, we keep the
potential as simple as possible so as to be able to equilibrate
the system at these temperatures.

\section{Results}

\begin{figure}
\includegraphics[angle=270,width=8.7cm]{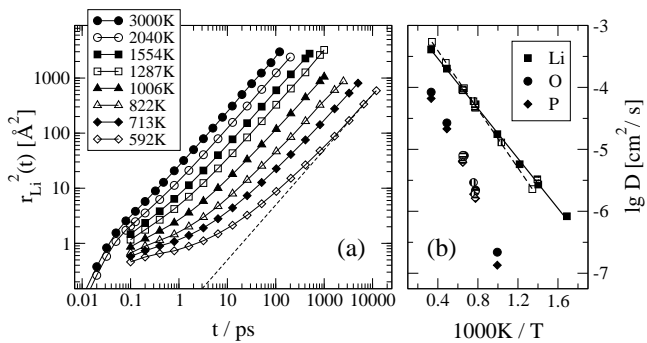}
\caption{(a): MSD of the lithium ions, $r^2_{Li}(t)$, in
non-crystalline LiPO$_3$ at the indicated temperatures. Dashed line:
$r^2_{Li}(t)\!=\!6D_{Li}t$ with
$D_{Li}\!=\!8.31\cdot10^{-7}\mathrm{cm^2/s}$. (b): Temperature
dependence of the diffusion constants $D_{Li}$, $D_{O}$ and $D_{P}$;
solid symbols: $N\!=\!800$, $\rho\!=\!2.15\mathrm{g/cm^3}$, open
symbols: $N\!=\!800$, $p\!=\!7.0\mathrm{GPa}$, vertical stripes:
$N\!=\!400$, $\rho\!=\!2.15\mathrm{g/cm^3}$, horizontal stripes:
$N\!=\!200$, $\rho\!=\!2.15\mathrm{g/cm^3}$, lines: Arrhenius fits to
the data for $N\!=\!800$.}\label{MSD}
\end{figure}

To study the dynamics of non-crystalline LiPO$_3$, we first
display the mean square displacement (MSD) of the lithium ions
\begin{equation}
r^2_{Li}(t)=\;\langle\;[r^j(t)]^2\,\rangle=\;\langle\;[|\,\vec{r}^{\;j}(t_0\!+\!t)-\vec{r}^{\;j}(t_0)|\:]^2\,\rangle
\end{equation}
in Fig.\ \ref{MSD}(a). Throughout this article, the brackets
$\langle\dots\rangle$ denote the average over the lithium ionic
subsystem, $r^{j}(t)$ is the displacement of lithium ion $j$
during a time interval $t$ and $\vec{r}^{\;j}(t_0)$ specifies its
position at the time $t_0$. At temperatures $T\!\leq\!2040\mathrm
K$, $r^2_{Li}(t)$ exhibits three characteristic time regimes:
ballistic motion $r^2_{Li}\!\propto\!t^2$ at short times
$t\!<\!0.1\mathrm{ps}$, sublinear diffusion due to back-and-forth
motions of the lithium ions at intermediate times and linear
diffusion $r^2_{Li}\!\propto\!t$ at long times. Upon cooling the
diffusion of the lithium ions slows down and the time window of
sublinear diffusion is extended. A qualitatively similar behavior
is found for the MSD of the oxygen and the phosphorus ions,
$r^2_{O}(t)$ and $r^2_{P}(t)$ (not shown).

The diffusion constants $D_{Li,\,O,\,P}$ extracted from the
long-time behavior of $r^2_{Li,\,O,\,P}(t)$ are compiled in Fig.\
\ref{MSD}(b). Evidently, the lithium diffusion decouples from the
oxygen and the phosphorus dynamics. While $D_{Li}$ follows an
Arrhenius law with activation energy $E_D\!=\!0.40\mathrm{eV}$,
the curves $\lg\,D_{O,P}(1/T)$ show a downward bending. In
particular, on the $10\mathrm{ns}$-timescale of our simulation, a
basically rigid phosphate glass matrix exists below
$T\!\approx\!1000\mathrm{K}$. Therefore, we refer to
$T_g\!=\!1000\mathrm{K}$ as the computer glass transition of
LiPO$_3$. For comparison, the diffusion constants resulting from
the calculations for $N\!=\!200$ and $N\!=\!400$ and from the
constant pressure simulations are also included in Fig.\ 1(b). We
find no indication that the diffusion constants depend on the
system size. On the other hand, the temperature dependence of
$D_{Li}$ in the constant pressure runs ($E_D\!=\!0.47\mathrm{eV}$)
is stronger than in the constant volume simulations.

\begin{figure}
\includegraphics[angle=270,width=6.5cm]{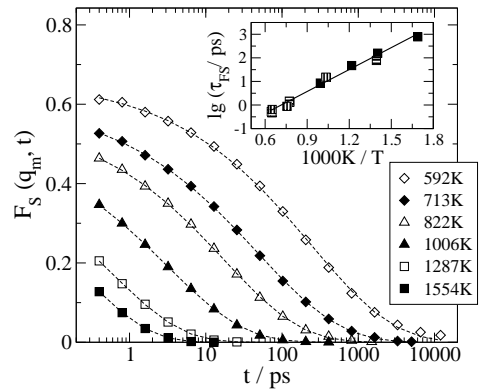}
\caption{Incoherent intermediate scattering function
$F_s(q_m\!=\!2.28\mathrm\AA^{-1},t)$ for the lithium ions in
non-crystalline LiPO$_3$ at the indicated temperatures ($N\!=\!800$,
$\rho\!=\!2.15\mathrm{g/cm^3}$). Inset: Temperature dependence of the
mean correlation time $\tau_{FS}$; solid symbols: $N\!=\!800$,
vertical stripes $N\!=\!400$, horizontal stripes: $N\!=\!200$, line:
Fit to an Arrhenius law for $N\!=\!800$. }\label{FS}
\end{figure}

Next, the incoherent intermediate scattering function $F_s(q,t)$
for the lithium ions is considered:
\begin{equation}
F_s(q,t)=\;<\!\cos\:[\,\vec{q}\,(\vec{r}^{\;j}(t_0\!+\!t)-\vec{r}^{\;j}(t_0))]\!>.
\end{equation}
In our case of an isotropic sample, this function only depends on the
absolute value, $q$, of the wave vector. To obtain information about
dynamical processes on the length scale of the Li-Li interatomic
distance, $q\!=\!q_m\!=\!2\pi/r_{LiLi}$ is used. Fig.\ \ref{FS} shows
$F_s(q_m,t)$ for the NEV ensemble ($N\!=\!800$). We see temperature
dependent, non-exponential decays. These findings can be quantified
by fitting the data to a Kohlrausch-Williams-Watts (KWW) function
\cite{KWW}, $A\,\exp[-(t/\tau)^{\beta}]$. Then, a stretching
parameter $\beta\!\approx\!0.42$ characterizes the non-exponentiality
at $T\!<\!T_g$. Above $T_g$, an increase to $\beta\!\approx\!0.54$ is
observed, however, the absence of a short-time plateau of
$F_s(q_m,t)$ leads to larger error bars in this temperature range. In
the inset of Fig.\ \ref{FS}, we display the mean time constants
$\tau_{FS}$ calculated from the fit parameters according to
$\tau_{FS}\!=\!(\tau/\beta)\Gamma(1/\beta)$ where $\Gamma(x)$ is the
$\Gamma$-function. Though there are some deviations, the temperature
dependence of $\tau_{FS}$ can still satisfactorily be described by an
Arrhenius law with activation energy $E_{FS}\!=\!0.62\mathrm{eV}$. In
the same way, $E_{FS}\!=\!0.67\pm0.03\mathrm{eV}$ is obtained from
the constant pressure simulations. Thus, in both cases, $E_{FS}$ is
significantly larger than $E_D$. In addition, Fig.\ 2 shows that the
time constants $\tau_{FS}$ do not depend on $N$ and, hence, finite
size effects are again absent. In the remainder of this article, we
analyze the data from constant volume simulations for $N\!=\!800$.

\begin{figure}
\includegraphics[angle=270,width=8cm]{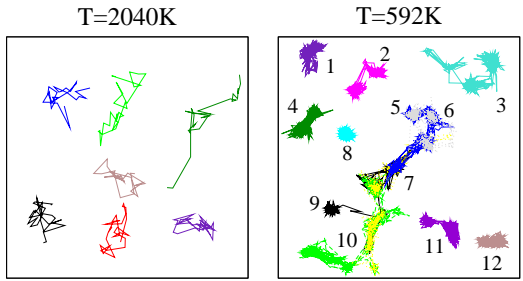}
\caption{Representative trajectories of the lithium ions in
non-crystalline LiPO$_3$. Left hand side: Trajectories at
$T\!=\!2040\mathrm K$ during a time interval $\Delta
t\!=\!3\mathrm{ps}$. Right hand side: Trajectories at
$T\!=\!592\mathrm K$ during a time interval $\Delta
t\!=\!100\mathrm{ps}$.}\label{tra}
\end{figure}

To study the mechanism of lithium dynamics, we first look at some
representative trajectories. Fig.\ \ref{tra} displays projections on
the $xy$-plane for trajectories during comparable time intervals at
$T\!=\!2040\mathrm K$ and $T\!=\!592\mathrm K$. While the mechanism
of the motion resembles liquid-like diffusion at the high
temperature, it is rather complex in the glass. Specifically, the
following features are obvious for $T\!=\!592\mathrm K$: (i) The
motion of most lithium ions can be decomposed into local vibrations
and jumps between adjacent sites. (ii) The lithium ions show
different mobilities. (iii) Back-and-forth dynamics involve distinct
lithium sites. (iv) The number of correlated back-and-forth jumps
varies. While numerous back-and-forth jumps are observed for ion 4,
only a few are found for ion 3. (v) Several ions (6-10) follow the
same preferential pathway. Consistent with the results of simulations
on alkali silicate glasses \cite{MD1,Manalang,Kieffer,Smith}, all
these features elucidate the complexity and the diversity of ionic
diffusion in glasses.

\begin{figure}
\includegraphics[angle=270,width=8.7cm]{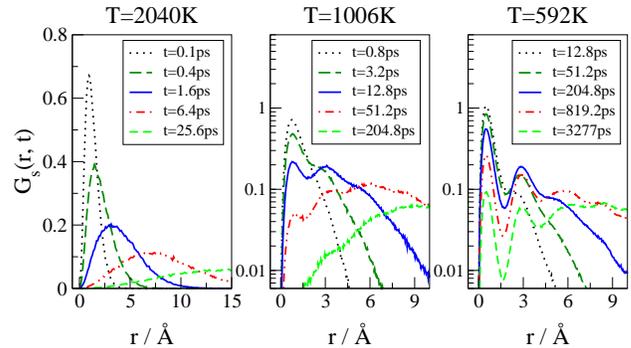}
\caption{Space- and time dependence of the self part of the van Hove
correlation function $G_s(r,t)$ for non-crystalline LiPO$_3$ at the
indicated temperatures.}\label{van}
\end{figure}

A statistical analysis of the mechanism of the motion is possible
based on the self part of the van Hove correlation function for
the lithium ions. It is given by \cite{Hansen}
\begin{equation}
G_s(r,t)=\langle\; \delta(\,r-r^j(t)\,)\;\rangle
\end{equation}
where $\delta(x)$ denotes the $\delta$-function. $G_s(r,t)$ measures
the probability that a lithium ion moves a distance $r$ in a time
interval $t$. Fig.\ \ref{van} shows $G_s(r,t)$ for three
characteristic temperatures. For $T\!=\!2040\mathrm K$, we observe a
single maximum that shifts to larger values of $r$ when $t$ is
increased. Though the findings resemble expectations for liquid-like
diffusion, an asymmetry of the curves for $t\!\approx\!1\mathrm{ps}$
indicates deviations. For $T\!\leq\!T_g$, $G_s(r,t)$ exhibits
oscillations that become more pronounced upon cooling. Specifically,
there are minima at $r\!\approx\!1.7 \mathrm \AA$ and
$r\!\approx\!4.2 \mathrm \AA$ together with a maximum at
$r\!\approx\!2.7\mathrm \AA\!\approx\!r_{LiLi}$. Thus, the glassy
network provides well defined lithium sites that are separated by
energy barriers $E\!>\!k_BT$ so that, to a good approximation, the
lithium ionic motion can be described as a sequence of hopping
processes, i.e., the timescale it takes to cross the barriers is much
shorter than the timescale the ions fluctuate about the sites. In
particular, it is possible to distinguish between ions that, after a
time $t$, (i) still reside at the same site
($r^j(t)\!\leq\!1.7\mathrm \AA$), (ii) have jumped to a next-neighbor
site ($1.7\mathrm \AA\!<r^j(t)\!<\!4.2\mathrm \AA$) and (iii) have
moved beyond an adjacent site ($r^j(t)\!\geq\!4.2\mathrm \AA$). These
findings for $G_s(r,t)$ agree with both the appearance of the
trajectories in Fig.\ \ref{tra} and the results of previous
simulations \cite{MD2,MD1,Smith,Bala,Habasaki2,Park,Horbach2}.

\begin{figure}
\includegraphics[angle=270,width=5.5cm]{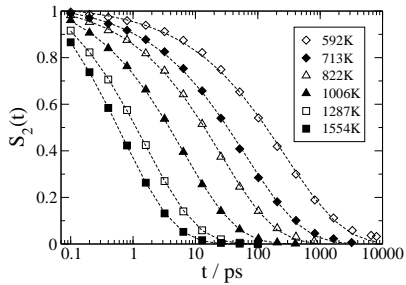}
\caption{Two-time correlation function $S_2(t)$ at the indicated
temperatures.}\label{S2}
\end{figure}

\begin{figure}[b]
\includegraphics[angle=270,width=5cm]{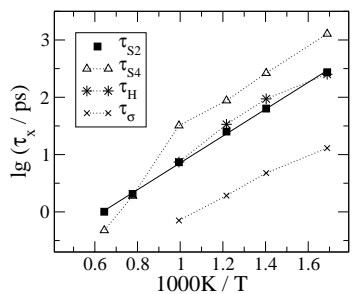}
\caption{Various time constants characterizing lithium ion dynamics
in non-crystalline LiPO$_3$, see text for details.}\label{tau}
\end{figure}

Next, we define the two-time correlation function
$S_2(t)\!=\!\langle s^{\,j}_2(t)\rangle$ where
$s^{\,j}_2(t)\!=\!1$ for $r^j(t)\!\leq\!1.7\mathrm \AA$ and
$s^{\,j}_2(t)\!=\!0$ otherwise. In our case of hopping motion,
$S_2(t)$ measures the probability that a lithium ion is still at
the initial site after a time $t$. Thus, it directly reflects the
depopulation of the initially occupied lithium sites. In Fig.\
\ref{S2}, we see that $S_2(t)$ decays non-exponentially. This
non-exponentiality can result from correlated back-and-forth jumps
between neighboring sites and/or from dynamical heterogeneities,
i.e., the presence of a distribution of jump rates. Based on an
appropriate three-time correlation function $S_3$, both
contributions will be quantified later in this article. The
average time it takes for the lithium ions to successfully escape
from their sites, i.e., the timescale of the depopulation of the
sites, can be characterized by the 1/e-decay time of $S_2(t)$,
$\tau_{S2}$. Fig.\ \ref{tau} shows that the temperature dependence
of $\tau_{S2}$ is well described by an Arrhenius law with
activation energy $E_S\!=\!0.44\mathrm{eV}$.

In the remainder of this article, the origin of the
non-exponential relaxation of the lithium ions in simulated
LiPO$_3$ glass is studied in detail by analyzing various
multi-time correlation functions. All these functions correlate
the positions of single lithium ions at subsequent times
$t_1\!<\!t_2\!<\dots$. To characterize the motion of ion $j$
between two times $t_{\alpha}\!<\!t_{\beta}$, we use the notation
$\vec{r}^{\;j}_{\alpha}\!\equiv\!\vec{r}^{\;j}(t_{\alpha})$,
$\vec{r}^{\;j}_{\alpha\beta}\!\equiv\!\vec{r}^{\;j}_{\beta}\!-\!
\vec{r}^{\;j}_{\alpha}$,
$r^{\;j}_{\alpha\beta}\!\equiv\!|\vec{r}^{\;j}_{\alpha\beta}|$ and
$t_{\alpha\beta}\!\equiv\!t_{\beta}-t_{\alpha}$.

\begin{figure}
\includegraphics[angle=270,width=8.7cm]{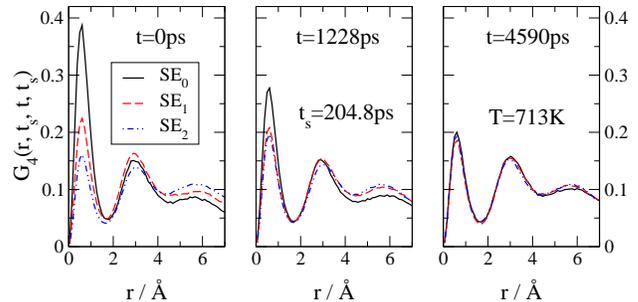}
\caption{Four-time van Hove correlation function
$G_4(r,t_{s},t,t_{s})$ for the lithium ions in LiPO$_3$ glass at
$T\!=\!713\mathrm K$. A filter time $t_s\!=\!204.8\mathrm{ps}$ was
used to select three subensembles of lithium ions SE$_0$, SE$_1$ and
SE$_2$, see text for details.}\label{G4}
\end{figure}

To show the existence of dynamical heterogeneities and to study
their lifetime, we define a ''four-time van Hove correlation
function''
\begin{equation}
G_4(r,t_{12}\!=\!t_s,t_{23},t_{34}\!=\!t_s)=\langle
a(r_{12})\,\delta(r-r^j_{34})\rangle.
\end{equation}
In this equation, the term $a(r_{12})$ simply means that three
subensembles are selected with respect to the displacement during the
time interval $t_{12}\!=\!t_s$: SE$_0$ consists of ions that are
still at the initial site ($r^j(t_{12})\!\leq\!1.7\mathrm \AA$),
SE$_1$ contains the ones that have jumped to an adjacent site ($1.7
\mathrm \AA\!<\!r^j(t_{12})\!<\!4.2\mathrm \AA$) and the ions of
SE$_2$ show $r^j(t_{12})\!\geq\!4.2\mathrm \AA$. Thus,
$G_4(r,t_s,t_{23},t_s)$ probes the van Hove correlation function
$G_s(r,t_s)$ of the respective subensembles starting at a time
$t_{23}$ after their selection. The results for
$t_s\!=\!204.8\mathrm{ps}$ and various $t_{23}$ at $T\!=\!713\mathrm
K$ are displayed in Fig.\ \ref{G4}. For $t_{23}\!=\!0$, the three
subensembles show clearly different $G_4$. In particular, the first
peak is highest for SE$_0$, indicating that ions that have not
escaped from their sites during $t_{12}$ move also less then the
average particle in the time interval $t_{34}$. In other words, slow
and fast ions are distinguishable and, hence, dynamical
heterogeneities exist \cite{12}. An extension of $t_{23}$, i.e., of
the delay between selection and detection, allows us to study the
lifetime of the dynamical heterogeneities. In Fig. \ref{G4}, we see
that the curves for the different subensembles approach each other
when $t_{23}$ is increased until they are nearly identical after a
time $t_{23}\!=\!4.6 \mathrm{ns}$. At this time
$t\!=\!4.6\mathrm{ns}$, the decay of $S_2(t)$ is just complete, cf.\
Fig.\ \ref{S2}. Thus, the information about the initial dynamical
states is lost when all ions have successfully escaped from their
sites, i.e., the dynamical heterogeneities are short-lived. A more
detailed analysis of the lifetime of the dynamical heterogeneities
will be presented below.

\begin{figure}
\includegraphics[angle=270,width=8.7cm]{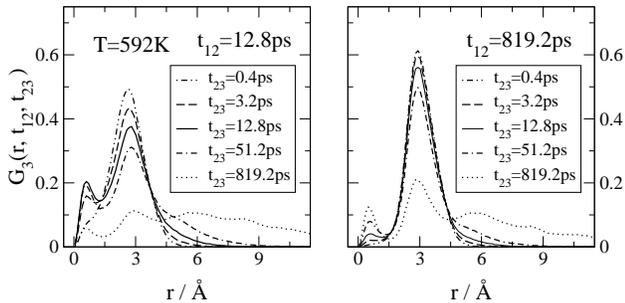}
\caption{Three-time van Hove correlation functions
$G_3(r,t_{12},t_{23})$ for the lithium ions in LiPO$_3$ glass at
$T\!=\!592\mathrm K$ }\label{G3}
\end{figure}

For a study of back-and-forth jumps, we focus on lithium ions that
have moved to a neighboring site during $t_{12}$ and map their
motion in the subsequent time interval $t_{23}$. This analysis is
possible based on the ''three-time van Hove correlation function''
\begin{equation}
G_3(r,t_{12},t_{23})=\langle\,d(r^j_{12})\,\delta(r-r^j_{13})\rangle
\end{equation}
where $d(r^j_{12})\!=\!1$ for $1.7\mathrm
\AA\!<\!r^j_{12}\!<\!4.2\mathrm \AA$ and $d(r^j_{12})\!=\!0$
otherwise. $G_3(r,t_{12},t_{23})$ measures
$G_s(r,t\!=\!t_{12}\!+t_{23})$ exclusively for ions that occupy a
next-neighbor site at the time $t_2$. $G_3(r,t_{12},t_{23}\!=\!0)$
is different from zero only in the range $1.7\mathrm
\AA\!<\!r\!<\!4.2\mathrm \AA$. For $t_{23}\!>\!0$, backward jumps
towards the initial positions $\vec{r}^{\;j}_1$ yield the
intensity $I_b$ at $r\!<\!1.7\mathrm \AA$, whereas forward jumps
contribute to the intensity $I_f$ at $r\!>\!4.2\mathrm \AA$.
Hence, when multiple jumps are negligible, i.e., for
$t_{23}\!<\!t_{12}$, the ratio $I_b/I_f$ provides information
whether the subsequent motion is backward or forward correlated.
In Fig.\ \ref{G3}, we display $G_3(r,t_{12},t_{23})$ at $T\!=\!592
\mathrm K$ for $t_{12}\!=\!12.8\mathrm{ps},\,819.2\mathrm{ps}$ and
various $t_{23}$. Nearly independent of $t_{23}\!<\!t_{12}$,
integration yields $I_b/I_f\!\approx\!15$ for
$t_{12}\!=\!12.8\mathrm{ps}$ and $I_b/I_f\!\approx\!0.3$ for
$t_{12}\!=\!819.2\mathrm{ps}$. While the former ratio indicates a
significant backward correlation, the latter is comparable to
$I_b/I_f\!\approx\!1/3$ as expected for a random-walk based on a
mean coordination number $z_{LiLi}\!\approx\!4$. For an
interpretation of these results, it is useful to recall the
existence of dynamical heterogeneities and to reinspect Fig.\
\ref{van}. It becomes clear that the population at the neighboring
site is still small after a time $t_{12}\!=\!12.8\mathrm{ps}$ and,
hence, it mainly results from the fast ions. Vice versa, ions with
a medium or a low mobility occupy an adjacent after a time
$t_{12}\!=\!819.2\mathrm{ps}$. Hence, the curves
$G_3(r,t_{12},t_{23})$ for $t_{12}\!=\!12.8\mathrm{ps}$ and
$t_{12}\!=\!819.2\mathrm{ps}$ reflect the behavior of fast and
non-fast ions, respectively. We conclude that a significantly
enhanced back-jump probability exists for the fast ions, but not
for the slow ions.

One may argue that large-amplitude vibrations can also lead to
$I_{b}/I_{f}\!\gg\!1$ for short $t_{12}$. However, one expects
that displacements due to vibrations are limited by
$r_{12}^j\!<\!r_{LiLi}$ and, hence, a significant contribution of
this type of motion would result in an asymmetry of
$G_3(r,t_{12},t_{23}\!\rightarrow\!0)$. Therefore, the nearly
symmetric shape of $G_3(r,t_{12},t_{23}\!\leq\!0.4\mathrm{ps})$
shows that large-amplitude vibrations do not affect our
conclusion. The finding that the back-jump probability depends on
the dynamical state is backed up by preliminary results of a more
detailed analysis. There, we explicitly identify the lithium sites
and calculate the back-jump probability $p_{\,b}$ as a function of
the waiting time at the initial site, $t_w$. Consistent with
results for LiSiO$_2$ \cite{Lammert}, we find that $p_{\,b}(t_w)$
significantly decreases with increasing waiting time. Thus, a high
backward correlation exists exclusively for ions that show a high
jump rate at the initial site.

The length scales of the dynamical heterogeneities and of the
correlated back-and-forth motions can be studied based on the
conditional probability functions $p\,(x_{23}|\,r_{12})$ and
$p\,(y_{23}|\,r_{12})$. These functions measure the probability to
find specific values $x^j_{23}\!=\!x_{23}$ and $y^j_{23}\!=\!y_{23}$,
respectively, provided the particle has moved a distance
$r^j_{12}\!=\!r_{12}$ in the first time interval $t_{12}$. Here,
$x^j_{23}$ is the projection of $\vec{r}^{\;j}_{23}$ on the direction
of the motion during $t_{12}$, i.e.,
$x^j_{23}\!=\!\vec{r}^{\;j}_{23}\cdot\vec{r}^{\;j}_{12}/r^j_{12}$ and
$y^j_{23}$ is the projection of $\vec{r}^{\;j}_{23}$ on an arbitrary
direction perpendicular to $\vec{r}^{\;j}_{12}$. Motivated by the
outcome of prior work \cite{MD2,Doliwa,Quian}, we focus on the first
moment $\overline{x}_{23}$ as well as on the second moments
$\sigma_x(r_{12})\!=\,<\![x^j_{23}\!-\!\overline{x}_{23}(r_{12})]^2(r^j_{12})\!>$
and
$\sigma_y(r_{12})\!=\,<\![y^j_{23}\!-\!\overline{y}_{23}(r_{12})]^2(r^j_{12})\!>$,
rather than analyzing the full probability functions.

\begin{figure}
\includegraphics[angle=270,width=8.7cm]{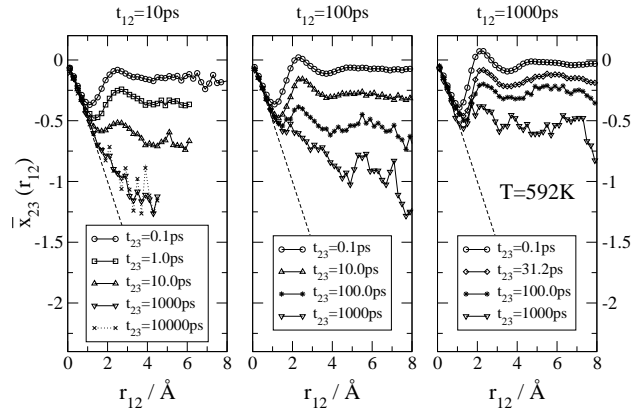}
\caption{First moment $\overline{x}_{23}(r_{12})$ of the conditional
probability function $p\,(x_{23}|\,r_{12})$ for LiPO$_3$ glass at
$T\!=\!592\mathrm K$. The time intervals $t_{12}$ and $t_{23}$ are
indicated.}\label{x12}
\end{figure}

For the first moment, $\overline{x}_{23}(r_{12})\!\equiv\!0$ will
result if the directions of the motions during $t_{12}$ and $t_{23}$
are uncorrelated \cite{Doliwa,Quian}. In contrast, if the subsequent
motion for a given $r_{12}$ is backward (forward) correlated, a
negative (positive) value of $\overline{x}_{23}(r_{12})$ will be
observed. In particular, it has been readily shown that
$\overline{x}_{23}(r_{12})\!=\!-(1/2)r_{12}$ is found for stochastic
dynamics in a harmonic potential \cite{MD2}. Fig.\ \ref{x12} shows
$\overline{x}_{23}(r_{12})$ for $T\!=\!592\mathrm K$. For all
$t_{12}$ and $t_{23}$, $\overline{x}_{23}(r_{12})$ follows the curve
$-(1/2)r_{12}$ up to $r_{12}\!=\!1.0\!-\!1.5\mathrm \AA$, indicating
that the lithium sites can be described by harmonic potentials on
this length scale. At larger $r_{12}$, $\overline{x}_{23}(r_{12})$
depends on both $t_{12}$ and $t_{23}$. For a given $t_{12}$,
$\overline{x}_{23}(r_{12})$ is more negative the lager the value of
$t_{23}$ until a saturation is reached at $t_{23}\!\approx\!1000
\mathrm{ps}$. This can be understood as follows: No significant
motion occurs during short time intervals $t_{23}$ so that
$x^j_{23}\!\approx\!0$. However, when $t_{23}$ is extended, more and
more lithium ions show a distinct displacement until the average
behavior of all ions with a given $r_{12}$ is observed for long
$t_{23}$. To study this average behavior we further discuss the
results for a long $t_{23}\!=\!1000\mathrm{ps}$. In this case,
$\overline{x}_{23}(r_{12})$ is basically constant at
$r_{12}\!>\!1\mathrm\AA$ for $t_{12}\!=\!1000\mathrm{ps}$, whereas,
for $t_{12}\!\leq\!100\mathrm{ps}$, a decrease over the whole
accessible $r_{12}$ range indicates that the backward correlation
increases even beyond the interatomic distance $r_{LiLi}$. A
comparison with Fig.\ \ref{van} shows that, for
$t_{12}\!\leq\!100\mathrm{ps}$, the findings in the range
$r_{LiLi}\!\leq\!r_{12}\!<\!8\mathrm\AA$, say, reflect the behavior
of fast ions, while an ion with an average mobility shows such
displacements during a time interval $t_{12}\!=\!1000\mathrm{ps}$.
Thus, the decrease of $\overline{x}_{23}(r_{12})$ at
$r_{12}\!\geq\!r_{LiLi}$ for short $t_{12}$ implies that the
correlated back-and-forth motions of fast ions may involve not only
one, but also several sites. On the other hand,
$\overline{x}_{23}\!\approx\!\mathrm{const.}$ at
$r_{12}\!>\!1\mathrm\AA$ for long $t_{12}$ again shows that the
non-fast ions perform local rattling motions, but no correlated
back-and-forth jumps between adjacent sites. We add that
$\overline{x}_{23}\!\approx\!-0.24r_{12}$ at $r_{12}\!<\!2\mathrm\AA
$ and $\overline{x}_{23}\!=\!const.$ at $r_{12}\!>\!2\mathrm\AA$ are
found for LiPO$_3$ melt at $T\!=\!2040\mathrm{K}$
($t_{12}\!<\!1\mathrm{ps}$, $t_{23}\!\geq\!1\mathrm{ps}$).
Qualitatively, these findings resemble the outcome of MD simulations
on supercooled liquids \cite{Doliwa,Quian}. There, the results were
interpreted as a signature of the cage effect describing that, though
all particles are mobile, every particle is temporarily captured in a
cage built up by its neighbors. Then, back-and-forth motions take
place within the cage, but not when the particles have left their
cages.

\begin{figure}
\includegraphics[angle=270,width=8.7cm]{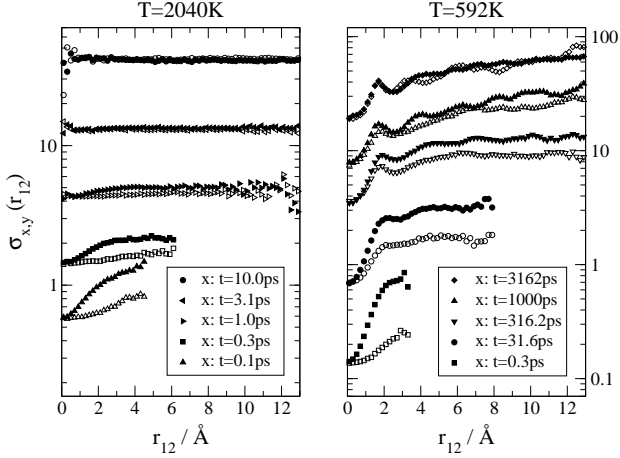}
\caption{Second moments $\sigma_x(r_{12})$ (solid symbols) and
$\sigma_y(r_{12})$ (open symbols) of the the conditional probability
functions $p\,(x_{23}|\,r_{12})$ and $p\,(y_{23}|\,r_{12})$,
respectively. Data for non-crystalline LiPO$_3$ at $T\!=\!2040\mathrm
K$ and $T\!=\!592\mathrm K$ are compared. The time intervals
$t\!=\!t_{12}\!=\!t_{23}$ are indicated.}\label{sigma}
\end{figure}

Now, we turn to the second moments. If all particles have equal
mobility, the displacements $r_{12}$ and $r_{23}$ are uncorrelated
so that $\sigma_{x,\,y}(r_{12})\!=\!\mathrm{const.}$. In contrast,
if dynamical heterogeneities exist, on average, slow and fast ions
show small and large $r_{12}$, respectively. Hence, ions with
small $r_{12}$ move less in $t_{23}$ than those with large
$r_{12}$ and $\sigma_{x,\,y}(r_{12})$ increases. Fig.\ \ref{sigma}
shows $\sigma_{x}(r_{12})$ and $\sigma_{y}(r_{12})$ for various
time intervals $t\!=\!t_{12}\!=\!t_{23}$ at two temperatures. For
$T\!=\!2040\mathrm K$, the second moments are equal and
independent of $r_{12}$ for $t\!>\!1\mathrm{ps}$, indicating the
absence of dynamical heterogeneities related to lithium diffusion.
In contrast, $\sigma_{x}(r_{12})$ and $\sigma_{y}(r_{12})$ for
$t\!\leq\!1\mathrm{ps}$ increase, though, with a different slope
and, thus there are highly anisotropic dynamical heterogeneities
on short timescales. Since
$r_{Li}(t\!<\!1\mathrm{ps})\!<\!r_{LiLi}$, cf.\ Fig.\ 1, we
attribute the latter dynamical heterogeneities to the rattling
motions within the local cages. For $T\!=\!592\mathrm K$, a strong
$r_{12}$ dependence of $\sigma_{x,y}$ is observed for all chosen
$t$. This finding again implies the existence of dynamical
heterogeneities attributed to lithium diffusion in the glass.
Since back jumps during $t_{23}$ result in $x_{23}^j\!\neq\!0$ and
$y_{23}\!\approx\!0$, i.e., the component perpendicular to the
direction of the forward jump is small, the dynamical
heterogeneities can be most directly studied based on
$\sigma_y(r_{12})$. The ratio
$\sigma_{y}(r_{LiLi})/\sigma_{y}(0)\!\gg\!1$ shows that the
heterogeneity results for the most part from different jump rates
at neighboring lithium sites. However, additional contributions on
a length scale of several Li-Li interatomic distances are
indicated by a further increase of $\sigma_{y}(r_{12})$ up to
$r_{12}\!\approx\!13\mathrm\AA$ for $t\!\geq\!1000\mathrm{ps}$.
Considering also the appearance of the lithium trajectories in
Fig.\ \ref{tra}, we suggest that preferred pathways of ion
migration connect several lithium sites.

The oscillatory behavior of $\overline{x}_{23}(r_{12})$ and
$\sigma_{x,\,y}(r_{12})$ allows us to study the energy landscape
governing the lithium dynamics. For $\overline{x}_{23}(r_{12})$,
one expects local minima (maxima) near positions where the slope
of the potential shows maxima (minima) \cite{MD2}. In Fig.\
\ref{x12}, we see that $\overline{x}_{23}(r_{12})$ shows a minimum
at $r_{12}\!\approx\!1.1\mathrm \AA$ and a maximum at
$r_{12}\!\approx\!2.3 \mathrm \AA$. Hence, there should be an
energy barrier at an average distance $r\!\approx\!1.7 \mathrm
\AA$. At $r_{12}\!\approx\!1.7 \mathrm \AA$, the curves
$\sigma_{x,\,y}(r_{12})$ exhibit local maxima. This is reasonable,
because ions residing at a saddle after $t_{12}$ should possess a
comparatively high mobility in the subsequent time interval. A
closer inspection of the data reveals that the maxima of
$\sigma_{x,\,y}(r_{12})$ shift to smaller $r_{12}$ when $t$ is
extended. As will be discussed later in this article, this finding
is consistent with the assumption that the Li-Li interaction
depends on the waiting time at a site.

\begin{figure}
\includegraphics[width=4.5cm]{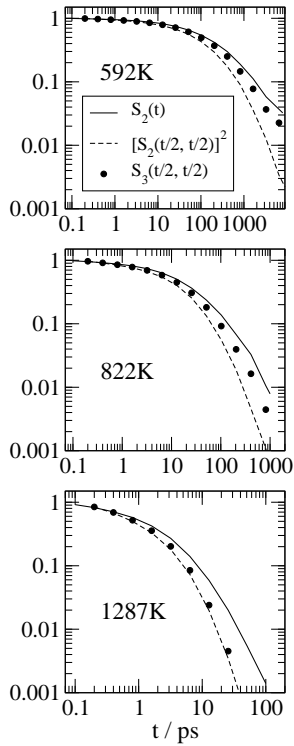}
\caption{ Three-time correlation functions $S_{3}(t/2,\,t/2)$ for
non-crystalline LiPO$_3$ at the indicated temperatures. The results
are compared to the expectations for purely heterogeneous dynamics,
$S_2(t)$, and for purely homogeneous dynamics,
$[S_2(t/2)]^2$.}\label{S3}
\end{figure}

To quantify the homogeneous and the heterogeneous contributions to
the non-exponential relaxation of the lithium ions we define the
three-time correlation function
\begin{equation}
S_{3}(t_{12}\!=\!t/2,\,t_{23}\!=\!t/2)=\;<\!s_2^j(t_{12})s_2^j(t_{23})\!>.
\end{equation}
For purely homogeneous and purely heterogeneous relaxation, such
three-time correlation functions can be expressed by the
corresponding two-time correlation functions
\cite{Heuer,Doliwa,Tracht,TrachtDR}. While
$S_{3}(t/2,t/2)\!=\![S_2(t/2)]^2$ is valid in the homogeneous
scenario, $S_{3}(t/2,t/2)\!=\!S_2(t)$ holds in the heterogeneous
scenario, cf.\ Appendix A. In Fig.\ \ref{S3}, $S_{3}(t/2,t/2)$ is
shown for three temperatures. For $T\!=\!1287K$,
$S_{3}(t/2,t/2)\!\approx\![S_2(t/2)]^2$ implies that the
relaxation is basically homogeneous. However, above $T_g$, hopping
dynamics is not a good approximation for the mechanism of the
motion and the phosphate matrix dynamics modifies the lithium
sites. Therefore, our findings do not indicate back-and-forth
\emph{jumps} between adjacent lithium sites, but rather they
reflect back-and-forth \emph{motions} that are affected by the
matrix dynamics and take place on various length scales. At
$T\!<\!T_g$, $S_{3}(t/2,t/2)$ clearly deviates from $[S_2(t/2)]^2$
and, hence, dynamical heterogeneities become increasingly
important upon cooling. Their relevance can be further analyzed
based on the ratio
\begin{equation}\label{HEQ}
H(t)=\frac{S_3(t/2,t/2)-[S_2(t/2)]^2}{S_2(t)-[S_2(t/2)]^2}.
\end{equation}
\begin{figure}
\includegraphics[angle=270,width=5cm]{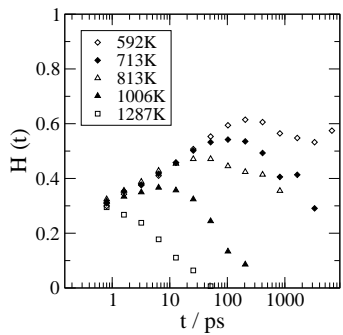}
\caption{$H(t)$ for non-crystalline LiPO$_3$ at the indicated
temperatures. The data were calculated from the results in Fig.\ 11
using Eq.\ \ref{HEQ}.}\label{H}
\end{figure}
It is easily seen that $H(t)\!\equiv \!1$ and $H(t)\!\equiv \!0$
hold for purely heterogeneous and purely homogeneous dynamics,
respectively. In Fig.\ \ref{H}, $H(t)$ is displayed for the
studied temperatures. Evidently, the non-exponentiality at low
temperatures results to a large extent from dynamical
heterogeneities, i.e., there is a broad distribution of jump
rates. For $T\!\leq\!1006K$, the curves $H(t)$ show a maximum at a
time $\tau_{H}$ that increases upon cooling. Inspecting the
temperature dependence of $\tau_{H}$ in Fig.\ \ref{tau},
$\tau_{H}(T)\!\approx\!\tau_{S2}(T)$ is found. Thus, the
distribution of jump rates contributes most to the
non-exponentiality when the lithium ions successfully escape from
their sites.

\begin{figure}
\includegraphics[angle=270,width=8.7cm]{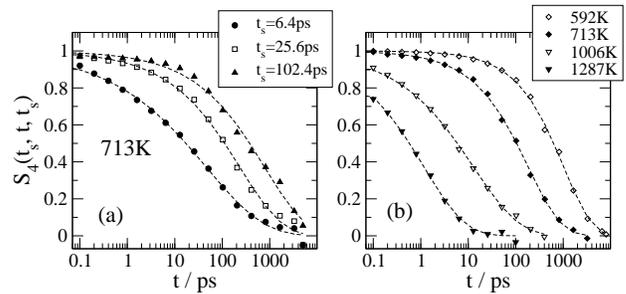}
\caption{Four-time correlation functions $S_4(t_s,t,t_s)$ for
non-crystalline LiPO$_3$ (points) and fits to a KWW function (dashed
lines). (a): Data at $T\!=\!713\mathrm K$ for the indicated filter
times $t_s$. (b): Temperature dependence of the curves for comparable
filter times, i.e., $S_2(t_s)\!=\!0.48\pm0.02$ ($592\mathrm{K}$:
$t_s\!=\!145.0\mathrm{ps}$, $713\mathrm{K}$:
$t_s\!=\!40.0\mathrm{ps}$, $1006\mathrm{K}$:
$t_s\!=\!4.0\mathrm{ps}$, $1287\mathrm{K}$:
$t_s\!=\!1.2\mathrm{ps}$).}\label{S4}
\end{figure}

The lifetime of the dynamical heterogeneities can be measured
based on the four-time correlation function
\begin{equation}
S_4(t_{12}\!=\!t_s,t,t_{34}\!=\!t_s)=\;<\!s_2^j(t_{12})s_2^j(t_{34})\!>.
\end{equation}
In this experiment, a dynamical filter selecting lithium ions that
are slow on the timescale $t_s$ \cite{slow} is applied during two
time intervals $t_{12}\!=\!t_s$ and $t_{34}\!=\!t_s$ separated by a
time $t$. Thus, $S_4(t_s,t,t_s)$ is given by the fraction of lithium
ions that are slow during $t_{12}$ \emph{and} a time $t$ later during
$t_{34}$. Consequently, the signal decreases when slow ions become
fast in the time interval $t$ and mapping $S_4(t)$ allows us to
measure the timescale of the exchange of the dynamical state. In
Fig.\ \ref{S4}, various normalized correlation functions
$S_4^{\,n}(t)$ are compiled. In panel (a), we show results for
different filter times $t_s$ at $T\!=\!713\mathrm K$ and, in panel
(b), we display data for comparable $t_s$ at various temperatures
($S_2(t_s)\!\approx\!0.48$). For the normalization, $S_4(t)$ was
fitted to a modified KWW function, $\Delta
S_4\exp[-(t/\tau)^{\beta}]\!+\!S_4^{\infty}$, and scaled according to
$S_4^{\,n}(t)\!=\!(S_4(t)\!-\!S_4^{\infty})/\Delta S_4$. Evidently,
$S_4^{\,n}(t)$ depends on both the filter time and the temperature
where all decays are non-exponential. The non-exponentiality
indicates that the exchange processes between slow and fast ions
cannot be described by a single time constant, but are rather
governed by a broad distribution of exchange times. Further, since
dynamically diverse subensembles are selected for different filter
times $t_s$, the dependence on this parameter implies that the
exchange time depends on the dynamical state itself. A comparison
with Fig.\ \ref{S2} reveals that the decays of $S_2(t)$ and $S_4(t)$
are complete at comparable times. In accordance with the
interpretation of $G_4$, this finding shows that the information
about the initial dynamical states is lost when all lithium ions have
successfully escaped from their sites. The mean time constants
$\tau_{S4}$, i.e., the mean exchange times, can be calculated from
the parameters of the KWW fits to $S_4(t)$. Revisiting Fig.\
\ref{tau} we see that $\tau_{S4}$ and $\tau_{S2}$ show a similar
temperature dependence at $T\!\leq\!T_g$. All these findings
consistently indicate that the exchange of the dynamical state in the
glass results from successful jumps among neighboring lithium sites.
Above $T_g$, a stronger temperature dependence of $\tau_{S4}$ is
observed. Hence, the modification of the lithium sites due to matrix
dynamics provides another channel for the exchange of the dynamical
state in the melt.

\section{Discussion}

In this article, we presented MD simulations performed to study the dynamics
of non-crystalline LiPO$_3$ in a temperature range where the
motions of the various atomic species strongly decouple, i.e., the
dynamics in the melt ($T\!>\!T_g$) and in the glass ($T\!<\!T_g$)
were compared. We found that the mechanism of the lithium motion
changes from liquid-like diffusion to hopping dynamics when the
temperature is decreased. Our main goal was to study the lithium
ionic jumps at $T\!<\!T_g$. Non-exponential two-time correlation
functions, such as the incoherent intermediate scattering
function, and sublinear diffusion over several orders of magnitude
in time reflect the complexity of the lithium motion in a rigid
phosphate glass matrix. For a comprehensive analysis of this
dynamical process, we calculated suitable multi-time correlation
functions. Three time-correlation functions showed that the
non-exponential relaxation of the lithium ions at $T\!<\!T_g$
results from both correlated back-and-forth jumps and the
existence of dynamical heterogeneities, i.e., the presence of a
distribution of jump rates. Measuring the lifetime of the
dynamical heterogeneities by means of four-time correlation
functions, we found that the non-uniformities of the motion are
short-lived. In what follows, we summarize and discuss our results
in more detail where the quantities that lead to the respective
conclusion are indicated in brackets.

The dynamical heterogeneities at $T\!<\!T_g$ show the following
features: (i) The distribution of jump rates contributes more to
the non-exponentiality the lower the temperature ($S_3$, $H$).
(ii) For a given temperature, its contribution is maximum when the
lithium ions successfully escape from their sites
($\tau_H/\tau_{S2}$). (iii) The exchange between the fast and the
slow ions of the distribution takes place on the same timescale as
the lithium jumps, i.e., the dynamical heterogeneities are
short-lived ($G_4$, $S_4$). In particular, both processes show a
similar temperature dependence ($\tau_{S4}/\tau_{S2}$). (iv) The
timescale of the exchange depends on the dynamical state itself
($S_4$). (v) The dynamical heterogeneities mainly result from
different jump rates at neighboring sites, but there are also
non-uniformities of the motion on a length scale
$r\!\geq\!10\mathrm\AA$ ($\sigma_{x,\,y}$). The back-and-forth
jumps of the lithium ions in the glass can be characterized as
follows: (i) Correlated back-and-forth jumps occur for the fast
ions of the distribution, but not for the slow ions, i.e., the
back-jump probability depends on the initial dynamical state
($G_3$, $\overline{x}_{23}$). (ii) For highly mobile ions, there
is a backward correlation even beyond the first neighbor shell
($\overline{x}_{23}$). Well above $T_g$, dynamical heterogeneities
and back-and-forth motions on the length scale $r_{LiLi}$ are
absent ($\overline{x}_{23}$, $\sigma_{x,\,y}$). However, due to
the temporary caging of the lithium ions by the neighboring
particles, anisotropic dynamical heterogeneities and a backward
correlation exist on short length- and timescales.

We conclude that the formation of well-defined lithium sites is
accompanied by the emergence of correlated back-and-forth jumps
and dynamical heterogeneities on a length scale
$r\!\geq\!r_{LiLi}$. At the lower temperatures, the
non-exponential depopulation of the lithium sites results to a
large extent from the dynamical heterogeneities and, hence, the
distribution of jump rates is broad. The short lifetime of the
dynamical heterogeneities implies that sites featuring fast and
slow lithium dynamics, respectively, are intimately mixed.
In particular, the absence of long-lived heterogeneities excludes a
micro-segregation of the glass into extended clusters in which the
lithium ions are mobile and immobile, respectively. Instead, we
suggest that the broad distribution of jump rates in LiPO$_3$
glass results from a diversity of the local glass structure at
neighboring sites, e.g., from varying coordination numbers. In
this basically random energy landscape, fast ion migration occurs
along pathways connecting low energy barriers. The presence of
such channels is not only indicated by the appearance of the
lithium trajectories, but also by the existence of long-range
dynamical heterogeneities and by the backward correlation beyond
the first neighbor shell for the fast ions.

Some of our findings are recovered when analyzing the dynamics
resulting from schematic models such as the random-barrier and the
random-energy model where the particles move in a
time-independent, external energy landscape. For example, the
particles migrate along preferential pathways in these models and
percolation approaches have been applied to describe the dynamics
at low temperatures \cite{Dyre,Baranovskii,Efros}. Moreover, the
back-jump probability depends on the dynamical state of a
particle, e.g., for the random-barrier model, many back-and-forth
jumps occur over low energy barriers, whereas slow ions cross high
barriers so that a backward motion is unlikely. However, other
features of the lithium dynamics in LiPO$_3$ glass are not
reproduced in schematic models. At variance with our findings,
cf.\ Fig.\ \ref{H}, a straightforward calculation shows that
$H(t\!\rightarrow\!0)\!\approx\!0.5$ is obtained for these models.
Furthermore, we observe that the first maximum of
$\sigma_{x,\,y}(r_{12})$ lies at $r\!>\!r_{LiLi}/2$ and shifts
towards shorter distances when $t$ is increased, cf.\ Fig.\
\ref{sigma}. This means that, in contrast to what is captured in
simple models, the energy barriers crossed by the fast lithium
ions of the distribution are located at larger distances than the
ones overcome by the slow ions. We speculate that simple hopping
models are suited to study the time-independent effects of the
glassy matrix on the lithium ionic diffusion, but they fail in
modelling the influence of the time-dependent Li-Li interactions.
For example, we want to demonstrate that our findings for the
position of the energy barrier can be explained when assuming that
the Li-Li interaction decreases with the waiting $t_w$ at a site
due to the adjustment of the neighboring lithium ions. For
simplicity, we consider a lithium ion that occupies a site at
$r\!=\!0$ and experiences a potential $V(r,t_w)\!=\!-\cos(\pi
r)\!+\!(1/2)a(t_w)\,r^2$. Here, the first and the second term are
meant to mimic the contribution from the glassy matrix and from
the repulsion due to other lithium ions ($a(t_w)\!\geq\!0$),
respectively. A Taylor expansion for $r\!\approx\!1$ shows that
the maximum of this potential lies at
$r_m(t_w)\!=\!1/(1-a(t_w)/\pi^2)$. Hence, for
$da(t_w)/dt_w\!<\!0$, this simple model yields a maximum at
$r_m(t_w)\!\geq\!1$ that is located at smaller distances the
larger the value of the waiting time as observed in our
simulations.

Comparing the results of MD simulation studies on phosphate and
silicate glasses several similar features of cation dynamics are
striking: (i) It is widely accepted that cation diffusion at
sufficiently low temperatures can be described as a sequence of
hopping processes
\cite{MD2,MD1,Smith,Bala,Habasaki2,Park,Horbach2}. (ii)
$F_s(q_m,t)$ consistently shows non-exponential decays that are
well described by KWW functions with stretching parameters
$\beta\!=\!0.42\!-\!0.47$ \cite{Horbach,MD1}. (iii) Dynamical
heterogeneities with a limited lifetime exist \cite{Habasaki} that
become more pronounced when the temperature is decreased
\cite{MD2}. (iv)  Fast ion transport along preferential pathways
takes place \cite{Jund,Sunyer}. (v) A high back-jump probability
exists for the fast ions of the distribution, but not for the slow
ions \cite{Lammert}. On the other hand, in previous work on sodium
silicate glasses \cite{Kob}, correlated back-and-forth jumps were
not observed at somewhat higher temperatures. However, since, a
high back-jump probability exists only for the fast ions at
sufficiently low temperatures, the percentage of ions showing
correlated back-and-forth jumps may have been too small to be
detected in the study by Sunyer et al.\ \cite{Kob}. Altogether,
though it is to be further clarified to which extent the findings
of MD simulations reflect the dynamics in a specific system at
lower temperatures, the similarity of the simulation results for
various interaction potentials suggests that the mechanism of
cation dynamics in various glassy ion conductors is highly
comparable.

Finally, we compare our results with experimental findings. First,
it can be noted that two-time correlation functions of lithium
dynamics obtained in experiments and simulations for LiPO$_3$
glass show a similar non-exponentiality. Specifically, the
stretching parameter $\beta\!\approx\!0.42$ found here compares
well to $\beta\!=\!0.33\!-\!0.54$ observed in electrical and
mechanical relaxation studies
\cite{Sidebottom,Green1,Green2,Martin} and in multi-dimensional
$^7$Li NMR experiments \cite{NMR}. Further, in these measurements,
an activation energy $E_a\!\approx\!0.68\mathrm{eV}$ was obtained
for the lithium dynamics. Comparing this value with the results of
our constant pressure simulations good agreement is found for
$E_{FS}\!\approx\!0.67\mathrm{eV}$, but significant differences
must be noted for $E_D\!=\!0.47\mathrm{eV}$. For the moment, it is
not clear where the difference between $E_{FS}$ and $E_D$ results
from. Since similar deviations were reported in MD simulation
studies of LiSiO$_2$ glass \cite{MD2}, one may speculate that they
are a consequence of the complexity of ion dynamics in glasses.

In addition, our findings are useful for the interpretation of
experimental results on the origin of the non-exponential
relaxation of the mobile ions in solid ion conductors.
Multi-dimensional $^{109}$Ag NMR experiments on glassy and
crystalline silver ion conductors \cite{Bunsen,PRL} showed that
the non-exponential silver ionic relaxation is due to the
existence of a broad distribution of jump rates rather than to
correlated back-and-forth jumps. On the other hand, the presence
of correlated back-and-forth motion is indicated by a frequency
dependence of the electrical conductivity $\sigma(\omega)$
\cite{Jonscher,Martin,Ingram,Elliot,Ngai,Funke,Roling}. We suggest
that this apparent discrepancy can be cleared up based on the
present results. Specifically, we found that, first, the back-jump
probability depends on the dynamical state, see also
\cite{Lammert}, and, second, the contribution of the dynamical
heterogeneities to the non-exponentiality increases upon cooling.
Since $\sigma(\omega)$ is dominated by the fast ions, we expect a
disproportional contribution from the correlated back-and-forth
jumps of ions in this dynamical state. In contrast,
multi-dimensional NMR experiments probe the behavior of non-fast
ions at low temperatures and, hence, if at all, back-and-forth
jumps should be of minor relevance.

In summary, various techniques have to be applied simultaneously
to obtain a complete picture of a complex dynamical process such
as ion dynamics in solids. In particular, a further comparison of
results obtained in MD simulations and in multi-dimensional NMR
experiments is very promising since both techniques allow one to
study a dynamical process on a microscopic level. For such a
comparison, the multi-time correlation functions $S_2$, $S_3$ and
$S_4$ can be particularly useful, because their information
content is similar to the one of the corresponding quantities in
NMR experiments.

\begin{acknowledgments}
The author thanks S.\ C.\ Glotzer, A.\ Heuer
and J.\ Kieffer for a generous grant of computer time and many
stimulating discussions. Founding by the Deutsche
Forschungsgemeinschaft (DFG) in the frame of the Emmy-Noether
Programm is gratefully acknowledged.
\end{acknowledgments}
\appendix

\section{Calculation of three-time correlation functions}

It is well established for supercooled liquids that, for purely
homogeneous and purely heterogeneous relaxation, three-time
correlation functions comparable to $S_3$ can be expressed by the
corresponding two-time correlation functions
\cite{Heuer,Doliwa,Tracht,TrachtDR}. Here,
$S_{3}(t/2,t/2)\!=\![S_2(t/2)]^2$ follows, because no selection is
possible for purely homogeneous dynamics. Hence, the dynamics
during $t_{12}$ and $t_{23}$, respectively, are statistically
independent so that the joint probability for an escape in neither
time interval can be written as a product. If this was not true a
dynamically diverse subensemble could be selected based on the
motion in the first time interval and, thus, dynamical
heterogeneities would exist. For purely heterogeneous dynamics,
the correlation functions result from a superposition of the
contributions of exponentially relaxing subensembles. For the
subensemble characterized by a correlation time $\tau_i$, one has
\begin{eqnarray*}
S_{2,i}(t)&\!=\!&\exp\left(-\frac{t/2+t/2}{\tau_i}\right)=S_{2,i}(t/2)S_{2,i}(t/2)\nonumber\\
&=&S_{3,i}(t/2,t/2)\nonumber.
\end{eqnarray*}
The second equals sign is due to the definition of the exponential
function, and the third results, because the relation for purely
homogeneous dynamics holds for the exponentially relaxing,
undecomposable subensembles. Since
$S_{3,i}(t/2,t/2)\!=\!S_{2,i}(t)$ is valid for all subensembles,
$S_{3}(t/2,t/2)\!=\!S_2(t)$ follows.


\begin{thebibliography}{99}
\bibitem{slow} Strictly speaking, the dynamical filter selects
lithium ions that are at the same site at the beginning and at the
end of the filter time $t_s$.
\end{thebibliography}
\end{document}